]Article

# Text Mining of Stocktwits Data for Predicting Stock Prices

**Mukul Jaggi \*, Priyanka Mandal, Shreya Narang, Usman Naseem and Matloob Khushi \***

1   School of Computer Science, The University of Sydney, Sydney NSW 2006, Australia; mkujaggi@gmail.com (M.J); pman7719@uni.sydney.edu.au (P.M); snar7221@uni.sydney.edu.au (S.N); engr.us-mannaseem87@gmail.com (U.N)
\*   Correspondence: matloob.khushi@sydney.edu.au (M.K.); mkujaggi@gmail.com (M.J.)

**Abstract:** Stock price prediction can be made more efficient by considering the price fluctuations and understanding people's sentiments. A limited number of models understand financial jargon or have labelled datasets concerning stock price change. To overcome this challenge, we introduced FinALBERT, an ALBERT based model trained to handle financial domain text classification tasks by labelling Stocktwits text data based on stock price change. We collected Stocktwits data for over ten years for 25 different companies, including the major five FAANG (Facebook, Amazon, Apple, Netflix, Google). These datasets were labelled with three labelling techniques based on stock price changes. Our proposed model FinALBERT is fine-tuned with these labels to achieve optimal results. We experimented with the labelled dataset by training it on traditional machine learning, BERT, and FinBERT models, which helped us understand how these labels behaved with different model architectures. Our labelling method's competitive advantage is that it can help analyse the historical data effectively, and the mathematical function can be easily customised to predict stock movement. The code and data are available from https://mkhushi.github.io/.

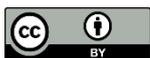

## 1. Introduction

Trillions of dollars are traded every day on financial markets [1,2]. The stock market is often deemed volatile and unpredictable, making it a challenge to maximise profit and hedging losses. There has been significant research invested in developing a valuable method in "beating the market". Among studying the historical information to identify the trends, in recent years, a lot of research is based on understanding people's sentiments using natural language processing (NLP) from social media platforms and its effect on the stock prices. Bollen et al. [3] was able to utilise an algorithm to determine Twitter users' sentiment towards the market and specific stocks that displayed a strong correlation between a stock's performance and the public's sentiment towards the asset.

This research's main objective is to identify the correlation between the changes in stock prices and the StockTwits data. In this research, the polarity of StockTwits data is dependent on the changes in a stock price, and this technique is incorporated by introducing a new data labelling method. For labelling the StockTwits dataset, historical data from Yahoo finance is extracted for the respective companies. Using this data, price change is calculated by comparing the Close price and Open price for the same day or previous day. Based on this calculation of price change, the StockTwits dataset is labelled. During the research, the StockTwits data was labelled in three different techniques: 1) Binary Classification: a technique in which StockTwits messages are labelled positive simply if the stock price increased and negative of the stock price decreased. 2) Percentage change two labels: with this technique StockTwits messages are labelled positive if the stock price increased more than 0.5% and negative if the stock price dropped less than 0.5%. 3) Percentage change three labels: like the previous method, the only difference is an addition of a third label neutral. If the price change is between –0.5% to +0.5% the StockTwits messages are labelled neutral. In this research, such labelling methods are experimented to understand which technique would generate the most optimal results.

Apart from labelling, a new model for predicting the Stock price movement is introduced. Currently, there exist various language models trained on a general corpus that are not effective in understanding the vocabulary related to the financial domain. To overcome this challenge, FinALBERT model is introduced. It is an ALBERT based model pre-trained on a financial corpus which would help the model to understand financial jargons. The pre-trained financial model is further fine-tuned on the labelled StockTwits data to predict the stock price movement.

## 2. Literature Review

In recent times, there is an increase in the use of feep learning and machine learning approaches on tex mining and financial domain [4-6]. In this section we review papers and experiments in the field of finance [7,8]. We also review labelling techniques natural language processing, machine learning and data extracted from a platform such as Twitter, and multiple ways of predicting stock performances based on it [8]. The most relevant reviews and results observed from the literature review are stated below.

Supriya et al. [9] and few others [8] explain binary classification, which is one of our labelling techniques categorises words as negative and positive, where positive terms were set to 1, and negative terms were set to 0. Percentage change as explained by Kordonis et al. [10], using a simple statistical function to estimate missing stock prices that were absent for weekends and other holidays when the market is closed using tweets from StockTwits and historical stock prices from Yahoo finance. It assumed that for all previously known value of stock price, "xprevious" and next known value to be "xnext", the value of missing value "y" is calculated as (xprevious+xnext)/2. Based on this percentage change classification on tweets was imposed by labelling tweets of the current day with change close price of the current day and previous day greater than +0.5 as positive, change greater than -0.5 as negative and change between [–0.5, +0.5] are labelled as neutral.

It was necessary to decide on the right set of models for our experiment with financial stock data, so we researched various models that already exist to understand the implementation, and the performance is compared. To understand further about transformer models, we looked into BERT: Bidirectional Encoder Representations from Transformers initiated by Devlin et al. [11], which talks about refining fine-tuning-based methods that have proved to obtain new state-of-the-art results on about 11 natural language processing tasks including GLUE (with 7.7% improvement) and SQuAD v2.0 (with 5.1 point improvement). The Liu et al. [12] study on BERT pre-training which was known to be expensive, significantly undertrained and used various datasets of various sizes and introduce RoBERTa to measure the significance of key hyperparameter and training data size that achieved state-of-the-art results on GLUE, RACE and SQuAD. The usage of financial stock Twitter data trained on a BERT model was conducted by Araci et al. [13] where FinBERT was introduced, which is a language model built out of BERT by pre-training it on a financial dataset and fine-tuning the same for sentimental analysis. Araci et al. [13] was able to achieve state-of-the-art on FiQA sentiment scoring and Financial PhraseBank, and the results were compared against ULMFit and ELMo for financial sentiment analysis.

GPU/TPU memory limitations are usually observed when there is a surge in model size and slower training period. To solve this problem, Lan et al. [14] introduced two-parameter reduction techniques to the original pre-trained BERT model, naming it to be ALBERT which uses fewer parameters only. The first technique uses the breakdown of large vocabulary embedding matrix into smaller ones to separate the size of hidden layers from that of vocabulary embedding, which will help in rising hidden size without increasing parameter size. The second technique explains a cross-layer parameter sharing to avoid the growth of parameter with the network's depth. As compared to BERT-large, the configuration of ALBERT-xxlarge achieves 30% parameter reduction, considerably improving the performance on SQuAD2 (+3.1%) and SST-2(+2.2%).

Kordonis et al. [10] introduced a system which collects previous financial stock connected tweets and used machine learning algorithms such as Support Vector Machine, achieving an accuracy of 0.8, and Naïve Bayes Bernoulli achieving an accuracy of 0.79 after categorising negative and positive sentiments. Word2vec and N-gram were utilised for analysing public sentiments in the tweets. Sentimental analysis, along with supervised machine learning principle, were applied to the tweets to analyse a correlation between stock market movement and sentiments observed in the tweets. Nabipour et al. [15] performs a comparative analysis of stock market trends prediction using machine learning and deep learning algorithms. This research compared nine machine learning methods such as Decision Tree, Random Forest, Adaptive Boosting, XGBoost, Support Vector Classifier (SVC), Naïve Bayes, K-Nearest Neighbours, Logistic Regression and Artificial Neural Network, and two deep learning methods such as Recurrent Neural Network and Long short-term memory. The results show that for the continuous data type, RNN and LSTM performed better than other prediction models with significant variance.

Das et al. [16] discourses how Twitter data can help in decision making such as stock market prediction to forecast prices of a company's stock by using RNNs Long Short-Term Memory (LSTM) which aided in extracting online stock data from various websites to predict future stock prices. Lu et al. [17] researched using CNN-BiLSTM-AM to predict the closing price of the stock for the next day. It involved neural networks (CNN), bi-directional long short-term Memory (BiLSTM) and attention mechanism (AM), wherein CNN was used to obtained features of the input data, BiLSTM uses the obtained feature data to predict the closing price of the next day. AM then used to seize the influence of feature positions on the closing price at multiple times in the past to improve the accuracy of the prediction. It was then concluded that CNN-BiLSTM-AM was more appropriate for the prediction of stock.

Ling et. al. [18], Zhao and Khushi [19] and Kim and Khushi [20] argue that it is best to predict the price of a financial instruct, and hence they propose to deal with this as a regression problem.

Pagolu et al. [21] found a robust association exists between the stock prices and public judgement based on tweets. So eventually, the study argued that if there's a greater number of positive tweets observed, then there is a chance that price might increase. This would, in turn, inspire people to invest in stocks.

Mishev et al. [22] performed more than 100 experiments using various datasets which were previously labelled by financial experts. Starting with a combination of machine learning and deep learning models lexicon-based approach, including word and sentence encoders and NLP transformers. It was concluded that even though the dataset used were comparatively small (with 2000 sentences), NLP transformers displayed excellent implementations compared to other assessed methods.

Models like BERT and FinBERT have a limitation of high training time and are computationally expensive due to many parameters. Also, the results of most state-of-the-art models do classification task based on sentiments, but for stock market prediction, sentiment analysis is not the only factor. The labelling techniques used in [9,10] have further experimented during the research. The proposed model overcomes the limitation by reducing the computational time compared to BERT and introducing another labelling technique that would help predict stock price movement.

## 3. Research Problems

*3.1. Research Aim and Objective*

This research intends to build a model based on historical stock prices and financial related messages to help predict the stock movement. The model would predict whether a Stocktwits message is positive, negative, or neutral, which would help understand stock price changes. Due to this, investors would better strategise their investment accordingly, i.e., whether to buy a stock or sell a stock. Accurate forecasting of the stock movement can decrease the investors' risk of investment and efficiently improve the return.

*3.2. Research Questions*

With chaotic market information, the stock market is continuously fluctuating. Investors face the equal opportunity of profit and loss in the stock market. A prediction made early on would always be advantageous in such a situation. Most of the current prediction models are either only based on sentiments or only on the stock prices. The data for a sentiment-based model is labelled based on sentiments, which is not the only factor needed for stock movement prediction. There exist only a few models based on a combination of user sentiments and the changes in stock prices. A model that generates temporally dependent prediction based on stock prices movement would help investors to decide to invest in a stock or not.

*3.3. Research Scope*

The research scope is to build a model that will classify and predict stock movement based on financial information within the StockTwits dataset. Also, a distinctive labelling method is followed where the historical stock price is considered while labelling the StockTwits dataset for predicting stock movement effectively. The transformer-based NLP model will be trained to recognise financial stock market jargon and fine-tuned to predict whether the message will positively or negatively affect the stock.

## 4. Materials and Methods

*4.1. Data Collection*

The study includes multiple datasets used for a different purpose. The StockTwits dataset used for training and testing, Yahoo Finance dataset for labelling while three external datasets (Reuters, AG News and bookcorpus) used for pre-training the proposed model.

4.1.1. StockTwits

StockTwits consist of data for 25 companies for over ten years. The companies included were Apple, Adobe, Tesla, Visa, etc. Each file contains the stock symbol, message, datetime, message id and user id for the respective messages. Sample rows for dataset is shown below in Table 1.

**Table 1.** StockTwits Data file.

| Symbol | Message | Datetime | User | Message_Id |
|---|---|---|---|---|
| TSLA | $TSLA trash | 2020-07-16T23:08:47Z | 3796654 | 2.28E+08 |
| TSLA | $TSLA https://www.tesmanian.com/blogs/tesmanian-blog/tesla-entering-greek-market ðŸ®ŽðŸš€ | 2020-07-16T23:06:01Z | 335497 | 2.28E+08 |
| TSLA | $TSLA what's happening here? Considerin selling out tomorrow! Convince me otherwise | 2020-07-16T23:04:50Z | 3572445 | 2.28E+08 |
| TSLA | $BB https://publishing.ninja/V4/page/10630/414/270/1 $TSLA | 2020-07-16T23:03:55Z | 1711636 | 228471610 |

4.1.2. Yahoo Finance

The StockTwits dataset is unlabelled, and there exists no other gold standard for labelled messages; thus, Yahoo Finance data was used for labelling. It consists of the historical stock prices for each day, and "yfinance" API was used to fetch the data. The data consists of the date, open price, high price, low price, close price, adjusted close price and volume (as shown in Table 2).

**Table 2.** Yahoo Finance data.

| Date | Open | High | Low | Close | Adj Close | Volume |
|---|---|---|---|---|---|---|
| 10/07/2020 | 279.2 | 309.784 | 275.202 | 308.93 | 308.93 | 1.17E+08 |
| 13/07/2020 | 331.8 | 358.998 | 294.222 | 299.412 | 299.412 | 1.95E+08 |
| 14/07/2020 | 311.2 | 318 | 286.2 | 303.36 | 303.36 | 1.17E+08 |
| 15/07/2020 | 308.6 | 310 | 291.4 | 309.202 | 309.202 | 81839000 |
| 16/07/2020 | 295.432 | 306.342 | 293.2 | 300.128 | 300.128 | 71504000 |

Apart from these datasets, three external datasets were used for pre-training our proposed model. These datasets include Reuters, AG_News and bookcorpus.

4.1.3. Reuters-21578 Dataset

Reuters - 21578 is a benchmark dataset for text categorisation, that appeared on Reuters newswire in 1987 [23]. This dataset has 22 files in total which are in SGML format. Files up till 21 consist of 1000 documents each while the 22$^{nd}$ file includes only 578 files. The documents in these files range over a variety of categories like places, people, money supply, exchanges, inventories, etc.

4.1.4. HuggingFace AG News

AG_News is a cluster of more than 1 million news articles, which were collected from more than 2000 sources of news. This dataset is used for research purposes in the field of data mining, information retrieval, etc. It consists of data belonging to four classes –

Business, Sports, Sci/Tech and World. For the study, only the data belonging to the Business class was selected because it was more related to the financial domain.

4.1.5. HuggingFace Book Corpus

Bookcorpus is a collection of text taken from Wikipedia. It consists of a total of 74004228 sentences aligning books to their movie releases. This dataset was used to train the model on regular English language so that an excellent English vocab is created. Due to hardware limitation for pre-training the proposed model, we have considered 10% of this dataset.

*4.2. Data Analysis*

4.2.1. Labelling Techniques

For labelling the messages in the StockTwits dataset three labelling techniques were used—Binary Classification (positive and negative), Percentage Change—2 labels (positive and negative) and Percentage Change—3 labels (positive, negative, and neutral). For each of these three techniques in total, the labelling was done using two ways—using same day open price—close price and using same day close price—previous day close price. Figure 1 shows hierarchy of different labelling techniques used.

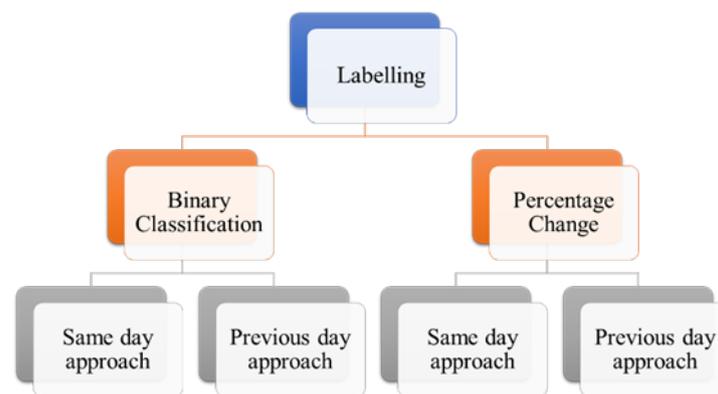

**Figure 1.** Labelling Approaches.

Since the Yahoo Finance data had missing data for weekends and public holidays, it was filled by comparing the previous date and next date or by using pandas dataframe functionality of date_range(). The values for the missing dates were filled using the logic cited in [3].

$$Xmean = \frac{X_{previous} + X_{next}}{2} \quad (1)$$

Where $X_{mean}$ is the missing value, $X_{previous}$ is the value of the field for the previously available date, and $X_{next}$ is the value of the field for the next available date.

The label for each date in Yahoo Finance data is calculated using the following equation -

For binary classification technique,
For same-day approach:

$$Label \begin{cases} 1 \ (positive), if \ close \ price > open \ price \\ 0 \ (negative), if \ close \ price < open \ price \end{cases} \quad (2)$$

For the previous day approach:

$$Label \begin{cases} 1 \ (positive), if \ current \ day \ close \ price > previous \ day \ close \ price \\ 0 \ (negative), if \ current \ day \ close \ price < previous \ day \ close \ price \end{cases} \quad (3)$$

For percentage change technique,
For same-day approach:

$$Percentage\ change = \langle\frac{Close_{sameday} - Open_{sameday}}{Open_{sameday}}\rangle * 100 \quad (4)$$

For the previous day approach:

$$Percentage\ change = \langle\frac{Close_{currentday} - Open_{previousday}}{Open_{previousday}}\rangle * 100 \quad (5)$$

The label is assigned according to the following criteria:

$$Percentage\ change = \langle\frac{Close_{currentday} - Open_{previousday}}{Open_{previousday}}\rangle * 100 \quad (6)$$

For each message in the StockTwits file, the label corresponding to its date in the Yahoo Finance data is allocated. Therefore, after labelling, the StockTwits file would contain the same content along with the label associated with each tweet.

4.2.2. Pre-Processing

Pre-processing of data is essential so that the machine can easily parse the data. Being a financial dataset with Twitter language, extra pre-processing was required apart from the regular pre-processing techniques. The messages contained not only words but also URLs, retweets, hashtags, emoticons, etc. The various techniques used were tokenisation, stop word, contractions, and repeated characters removal, eliminating mentions, dropping retweets, case folding, demojization, hashtag segmentation and URL removal.

4.2.3. Data Characteristics

StockTwits dataset has nearly 6.4 million records spread across ten years for all the 25 companies. For our experimentation, we limited this dataset to include only FAANG companies. FAANG includes the top five famous companies in the world—Facebook, Apple, Amazon, Netflix, and Google. The total number of tweets per company are as shown in Figure 2.

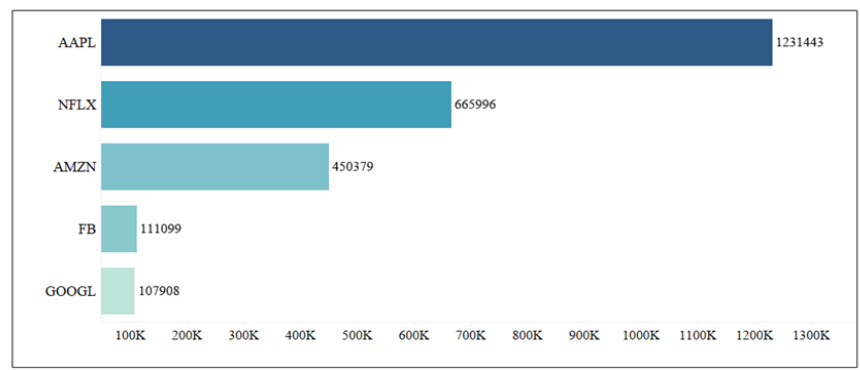

**Figure 2.** Company-wise FAANG dataset.

The total number of records in the FAANG dataset for the one-year duration is 514,320. The data characteristics for it using all three labelling techniques on one-year data is as shown in Figure 3.

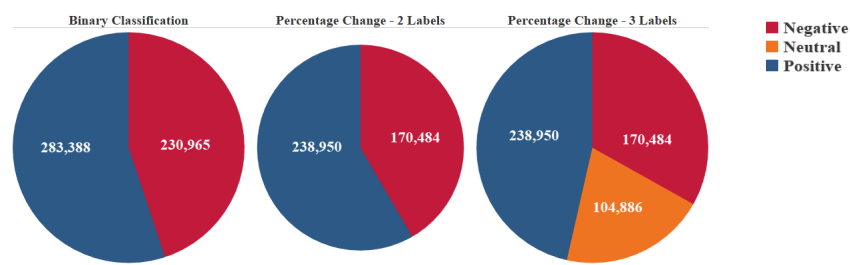

**Figure 3.** 1 Year FAANG data.

While using binary classification, 55% of the tweets are labelled as positive, and 45% of the tweets are labelled as negative. Using percentage change with two labels, 58.3% of tweets are positive, and 41.7% tweets are negative. When using percentage change with three labels, 46.5% tweets are positive, 33.1% tweets are negative, and the 20.4% tweets are neutral.

The total number of records in the FAANG dataset for two years duration is 1,021,417. The data characteristics for it using all three labelling techniques on two years data is as shown in Figure 4.

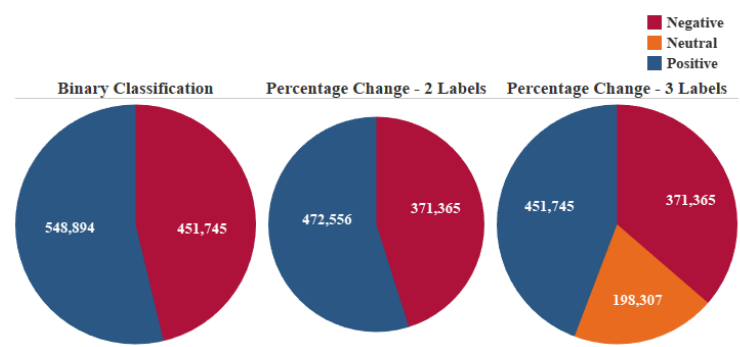

**Figure 4.** 2 Year FAANG data.

When using binary classification, 54.8% of the tweets are labelled as positive, and 45.2% of the tweets are labelled as negative. Using percentage change with two labels, 56% of tweets are positive, and 44% tweets are negative. Using percentage change with three labels, 44.2% of tweets are positive, 36.4% tweets are negative, and 19.4% tweets are neutral.

*4.3. Methods*

We experimented with three types of models – Transformer based Models, Traditional Machine Learning Models and Word Embeddings with Neural Networks.

**Traditional machine learning methods**

The following models were run on FAANG dataset on all the labelling techniques. The features of the tweets were extracted using TF-IDF Vectorizer and Count Vectorizer in a Pipeline fashion. The dataset was split into the training set and testing set using 90:10 ratio. The models were implemented using Grid Search, both with and without 5-fold cross-validation. The models were trained using modules of sklearn library.

4.3.1. Naïve Bayes

It is a supervised machine learning algorithm, which assumes that feature pairs are independent and is based on Bayes' theorem. It works on the assumption that all variables present in the dataset are not correlated with each other rather are Naïve. It has the most straightforward implementation, works well for large datasets and dataset with categorical data [24] Additionally, the training time for this model is extremely fast when compared to other traditional models.

4.3.2. Random Forest Classifier

It is based on multiple individual decision trees, in an ensemble way. This means that it considers prediction results from individual trees to decide the outcome, which leads to better performance. This is better than the prediction result of any individual tree alone.

4.3.3. Gradient Boosting Regressor

This is another Ensemble technique (boosting), in which predictors are made sequentially, and not independently. Each predictor learns from the mistakes made by the previous predictor. Hence, less time and iterations are needed to get the real predictions.

4.3.4. Logistic Regression

A statistical model, which is used when there is a dependent (target) variable. Firstly, linear regression is applied to fitting the data. Then, a logistic function is applied for predicting the probabilities of various classes of data. These probabilities are converted to binary form by using a sigmoid function, which helps make actual predictions [25]. It's less inclined to overfitting the model in a low dimensional dataset [26].

4.3.5. XGBoost

eXtreme Gradient Boosting is a supervised machine learning algorithm. It's based on decision trees and uses a gradient boosting framework. It also works well with tabular or structured data. XGBoost gives good performance due to system optimisation and algorithmic enhancements.

4.3.6. Transformer based models

To implement transformer-based models, a dedicated GPU environment is needed. For this purpose, Google Colab GPU with enabled CUDA was used for fine-tuning the models. CUDA (Compute Unified Device Architecture) is a parallel computing platform which speeds up the computation of applications running on it.

4.3.7. BERT

BERT, released by Google, is based on Transformer architecture, caused a stir in the NLP field as it showed state-of-the-art results on various NLP tasks. The innovation had a deeper sense of language context and flow as it was bidirectionally trained, as compared to the older neural network architectures. For implementing BERT, HuggingFace's transformer library was used, which contains various classes for BERT like BertTokenizer, BertForSequenceClassification, BertForTokenClassification, etc. For fine-tuning BERT, we used BertTokenizer and BertForSequenceClassification. Out of the many pre-trained BERT models available, we used bert-base-uncased, which consists of 12 layers, 768 hidden layers and 110M parameters. The maximum sequence length of tweets is 160 as this is the limit for Twitter tweets too. BertForSequenceClassification was used for fine-tuning. It has the same architecture as BERT along with an additional linear layer for classification. For model optimisation, AdamW optimiser was used.

4.3.8. FinBERT

Araci et al. [13] implemented FinBERT on Financial Phrasebank dataset, which contains labels as a string instead of numbers. Hence, the first step was to convert the labels in the StockTwits files from 1, 0 and -1 to positive, neutral and negative. The StockTwits dataset was split into training, testing and validation CSV files to be used in the implementation ahead. FinBERT language model was downloaded from the author's GitHub repository. Bert-base-uncased tokeniser and pre-built language models were used to fine-tune FinBERT on our dataset. FinBERT implementation was done using binary classification or percentage change – 2 labels techniques.

4.3.9. FinALBERT (Proposed Model Implementation)

Model Architecture: Figure 5 below shows the model architecture for FinALBERT. FinALBERT model includes two main steps the creation of the pre-trained model and fine-tuning on the pre-trained model to determine the model results. For pre-training with the FinALBERT model, we used three external datasets to build the model vocabulary. For fine-tuning the pre-trained model, the StockTwits dataset for each of the company was initially pre-processed and then labelled based on the three techniques mentioned earlier. After fine-tuning the model, the model results were evaluated, and labels were predicted for individual companies.

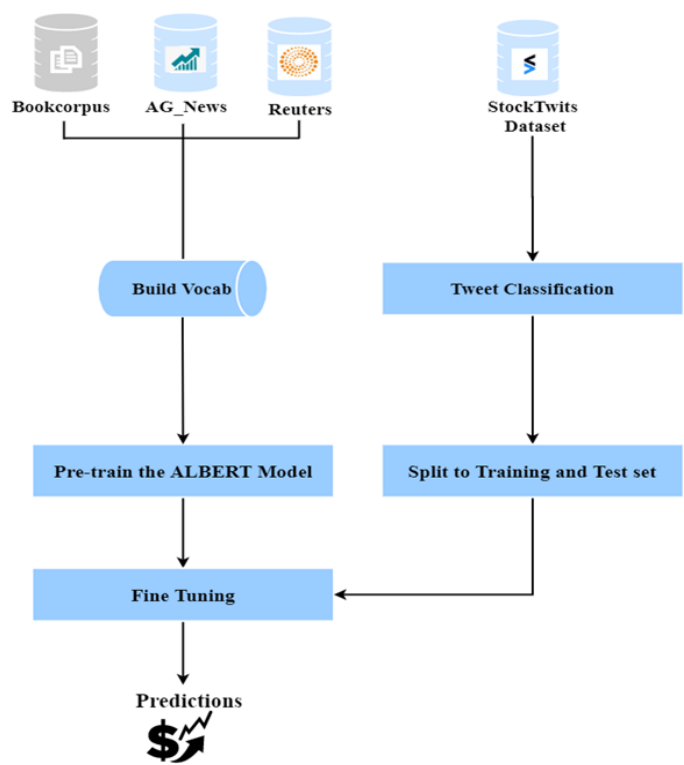

**Figure 5.** Model Architecture.

4.3.10. FinALBERT model Pre-training

The pre-training for FinALBERT model includes three steps: generating a vocab file, constructing the data in the format required for pre-training and the final pre-training the model step based on the model vocab and data file created in the above steps.

**Data creation for Pre-training the model:** To create the pre-trained FinALBERT model, we combined the three datasets: Reuters-21578, AG News and 10% of the Bookcorpus data. Before combining these datasets, they were individually pre-processed using the techniques mentioned earlier. The final data file created consisted of 7,430,280 sentences, and the file size was 450 MB.

**Building Model Vocab:** we used the SentencePiece tokeniser to generate the model vocab. SentencePiece is the same library used in ALBERT to build the model vocab. It is a text tokeniser which is unsupervised in which the size of the vocabulary is pre-determined. By using this library, the tweets are split into tokens by implementing subword units which include byte-pair encoding. Additionally, using this library is beneficial because it has no logic, which is language-dependent, and it considers the sentences as Unicode character sequences [27]. To generate the sentence piece model vocab file, we require the data in a text file with a space between individual sentences (as shown above) Also, the vocab size is given as input which we selected as 30k.

**Creation of Pre-training Data:** From the original ALBERT GitHub repository, we used the file create_pretraining_data.py. The input for this file is the combined data corpus created and the SentencePiece vocab file. Based on these input files, this file creates the training input file required for pre-training the ALBERT model. The sentences are mapped to corresponding input_ids, input_masks, segment_ids and masking positions, which are then used for model training. The output of this file is a binary file in which the data is in the format needed for pre-training the ALBERT model.

**Pre-training FinALBERT:** From the original ALBERT GitHub repository, we used the file run_pretraining.py to build the pre-trained model. Before executing this file, we needed to create another file known as albert_config.json. This file consists of the ALBERT model architecture and defines the size of embeddings, number of hidden layers and the number of attention heads.

Apart from the config file, the binary file created above is another input for run_pretraining.py. Other parameter inputs include train_batch_size equal to 256, max_seq_length which was set as 512, learning_rate as .00176 and num_train_steps as 10000. In the original configuration of albert-base model, the batch_size was 4096 and num_train_steps were kept as 1,25,000. 125k are the number of training steps for the ALBERT model, which help in controlling the data throughput [14]. The output of the run_pretraining.py is the pre-trained model stored in the output directory defined.

4.3.11. FinALBERT Fine-tuning

The next step was the fine-tuning of pre-trained FinALBERT model by using the StockTwits dataset. After loading, the dataset was split into training and test sets using the train_test_split of sklearn library. The data was split randomly as 90% training and 10% testing set. Once the datasets are created, the next step was loading the Albert models. For this, we used the HuggingFace transformers library from which we used AlbertTokenizer and AlbertForSequenceClassification. In the AlbertTokenizer was used another function called as encode_plus, which is used to tokenise and prepare the tweets for the model in the form of sequence pairs based on the pre-trained model vocabulary. The data is prepared by converting to input_ids, attentions_masks and adding the required special tokens. The data is converted into TensorFlow format required for model training. AlbertForSequenceClassification is used to build the model architecture, and load_tf_weights_in_albert is used to load the pre-trained FinALBERT model.

We experimented with different hyperparameter values of learning_rate (set as 1e-05, 2e-05 up to 5e-5) and different batch_sizes (8, 16, 32 and 64). ALBERT model is sensitive to the combinations of these hyperparameter values, and not every combination gave optimal results. We achieved the best model results by setting the learning rate as 1e-05, batch size as 32 and using the AdamW optimiser.

4.3.12. Prediction

For performing predictions, last two weeks of StockTwits data for each company is filtered and loaded as a pandas dataframe, separately. The fine-tuned FinALBERT model and tokeniser are loaded using AlbertForSequenceClassification and AlbertTokenizer, respectively. The data is prepared for input to the model in the same way as done during the fine-tuning stage. The data in the form of input_ids and attention masks are passed to the model, and softmax activation function is applied to the model output. Softmax function returns the label probabilities. Argmax function is applied to these probabilities to get the real predicted labels. Additionally, a threshold is set for the precision value of the positively predicted labels. If the precision values are more than the threshold, the model output is to invest in the stock else avoid investing.

**Word embeddings with neural network**

4.3.13. BiLSTM

LSTM is an RNN based architecture which is widely used in sequential learning problems. In Bidirectional LSTM there are two LSTM layers: one layer takes the input in a forward direction, with the second layer takes the input sequence in a backward direction; by this we can use past and future features together for better understanding of the model [28].

4.3.14. Attention

It is a mechanism which is implemented over a sequential language model. In language models, the focus is on the entire sequence of data, which might lead to information loss since important words are not focused. Attention mechanism selectively focuses on a sequence of input and calculates attention in context to the sequence. It calculates the softmax of each word (token) in the sequence according to its context, and the attention is focused on the word with the highest score. There are multiple attention mechanisms to calculate the attention score such as additive, content-based and dot-product etc. [29].

4.3.15. FinALBERT + BiLSTM +Attention

To experiment with the FinALBERT model, we extracted the word embeddings from the fine-tuned model and used these word-embeddings as the input for BiLSTM model. The word vectors were extracted by using the "albert-base-v2" tokeniser. Attention layer was implemented over the BiLSTM layer.

4.3.16. BERT Embeddings with CNN

For implementing CNN with BERT embeddings, BERT pre-trained model with bert-base-uncased vocabulary is used. Children() function is applied to the model to fetch all layers of the model, of which the last classification layer is extracted using indexing. Tensor representations of the model parameters, which are an extracted result in the embedding matrix. This is passed as weights to the embedding layer of CNN model, which contains three convolutional layers. Hence, each sentence is embedded using the matrix extracted from BERT.

4.3.17. Word2vec with CNN

For implementing the CNN model using Word2Vec embeddings, the Google news pre-trained Word2Vec model was used. This model has been trained on more than 100 billion words extracted from Google news data and includes word vectors for approximately 3 million phrases and words [30]. After extracting the word embeddings, the embedding dimension of training was passed as weights to the embedding layer of CNN model. On top of the input layer, convolutions were added to get the output of this model. Additionally, to get the final model results, the CNN layer output was passed through the Sigmoid function to get the prediction results in binary format.

## 5. Results

We performed various experimentations on the StockTwits FAANG dataset. With more than 6.4 million data samples, the baseline models, traditional machine learning models, and proposed models were trained for all set of possibilities to extract results. All the experimentations were done for three labelling techniques.

All the experimentations were done on FAANG data with 1 and 2 years of the subset. Result for different labelling methods trained on various baselines models are as follows: Table 3 shows all the results for Binary Classification with 1 and 2 years of data. Naïve Bayes performs best for one-year data, whereas, for two-year data, BERT performs best with a macro average F1 score of 0.59.

Boldface in each of the below table indicates the highest macro F1 score and accuracy among each of the labelling method for different year configuration.

**Table 3.** Results for Binary classified labels.

| | Binary Classification | | | |
|---|---|---|---|---|
| **Model** | 1 Year | | 2 Year | |
| | F1 Score-Macro | Accuracy | F1 Score-Macro | Accuracy |
| **Naive Bayes** | **0.58** | **59%** | 0.58 | 59% |
| **Gradient Boosting** | 0.37 | 55% | 0.36 | 54% |
| **Logistic regression** | 0.55 | 57% | 0.53 | 57% |
| **Random Forest** | 0.54 | 55% | 0.54 | 54% |
| **BERT** | 0.54 | 58% | **0.59** | **61%** |
| **FinBERT** | 0.54 | 54% | 0.54 | 54% |
| **FinALBERT** | 0.35 | 55% | 0.32 | 46% |
| **CNN with Word2Vec** | 0.40 | 47% | 0.41 | 54% |
| **CNN with BERT embeddings** | 0.35 | 55% | 0.35 | 54% |
| **Word2Vec with BiLSTM and Attention** | 0.56 | 57% | 0.52 | 57% |
| **Fasttext with BiLSTM and Attention** | 0.54 | 58% | 0.54 | 57% |
| **FinALBERT with BiLSTM and Attention** | 0.53 | 57% | 0.55 | 56% |

Table 4 shows all the results for Percentage change with two labels (Positive and Negative) with 1 and 2 years of data. BERT performs best for one-year data as well as for two-year data with a macro average F1 score of 0.59 and 0.60, respectively.

**Table 4.** Results for Percentage change labels (Positive and Negative).

| Model | Percentage Change with Positive and Negative Labels | | | |
|---|---|---|---|---|
| | 1 Year | | 2 Year | |
| | F1 Score-Macro | Accuracy | F1 Score-Macro | Accuracy |
| **Naive Bayes** | 0.58 | 62% | 0.59 | 61% |
| **Gradient Boosting** | 0.37 | 58% | 0.37 | 55% |
| **Logistic regression** | 0.49 | 60% | 0.53 | 58% |
| **Random Forest** | 0.55 | 56% | 0.55 | 55% |
| **BERT** | **0.59** | **64%** | **0.60** | **62%** |
| **FinBERT** | 0.54 | 54% | 0.53 | 53% |
| **FinALBERT** | 0.51 | 60% | 0.35 | 55% |
| **CNN with Word2Vec** | 0.42 | 59% | 0.46 | 56% |
| **CNN with BERT embeddings** | 0.38 | 59% | 0.35 | 55% |
| **Word2Vec with BiLSTM and Attention** | 0.52 | 60% | 0.56 | 59% |
| **Fasttext with BiLSTM and Attention** | 0.54 | 61% | 0.54 | 59% |
| **FinALBERT with BiLSTM and Attention** | 0.50 | 60% | 0.55 | 59% |

Table 5 depicts the results for Percentage change with three labels (Positive, Neutral, and Negative) with 1 and 2 years of data. The results generated are the lowest compared to other labelling techniques. Naïve Bayes performs better than any other model with a macro average F1 score of 0.41 on both years' dataset.

**Table 5.** Results for Percentage change (Positive, Neutral, and Negative).

| Model | Percentage Change with Negative, Neutral and Positive Labels | | | |
|---|---|---|---|---|
| | 1 Year | | 2 Year | |
| | F1 Score-Macro | Accuracy | F1 Score-Macro | Accuracy |
| **Naive Bayes** | **0.41** | **49%** | **0.41** | **49%** |
| **Gradient Boosting** | 0.11 | 20% | 0.1 | 19% |
| **Logistic regression** | 0.3 | 48% | 0.32 | 47% |
| **Random Forest** | 0.33 | 33% | 0.34 | 34% |
| **BERT** | 0.33 | 48% | 0.37 | 50% |
| **FinBERT** | 0.32 | 33% | 0.36 | 36% |
| **FinALBERT** | 0.29 | 47% | 0.32 | 46% |
| **CNN with Word2Vec** | 0.26 | 42% | 0.26 | 34% |
| **CNN with BERT embeddings** | 0.17 | 23% | 0.2 | 44% |
| **Word2Vec with BiLSTM and Attention** | 0.21 | 46% | 0.11 | 20% |
| **Fasttext with BiLSTM and Attention** | 0.21 | 47% | 0.2 | 44% |
| **FinALBERT with BiLSTM and Attention** | 0.21 | 46% | 0.21 | 44% |

Proposed model FinALBERT performs moderately with all three labelling techniques with the macro F1 score ranging between 0.29–0.51. The highest score achieved was with the percentage change method two labels. Percentage change with two labels has consistently performed better than the percentage change three labels. The binary classification has persistently generated similar results to percentage change two labels. BERT outperforms in 1-year data, whereas Naïve Bayes has achieved the highest F1 score in 2 years FAANG data.

Traditional machine learning models perform average with varying methods of labelling. Naïve Bayes, Logistic Regression and Random forest were trained using cross-validation (k-fold). Tables 6 and 7 below show the results for four different machine learning traditional methods.

**Table 6.** Traditional machine learning model results with K-fold (1 year).

| Model | Binary Classification | Percentage Change-2 Labels | Percentage Change-3 Labels |
|---|---|---|---|
| | | 1 Year | |
| | F1 Score-Macro | F1 Score-Macro | F1 Score-Macro |
| **Naïve Bayes** | 0.57 | **0.58** | 0.41 |
| Logistic Regression | 0.51 | 0.5 | 0.31 |
| Random Forest | 0.54 | 0.55 | 0.34 |
| Gradient Boosting | 0.34 | 0.38 | 0.11 |

**Table 7.** Traditional machine learning models results with K-fold (2 years).

| Model | Binary Classification | Percentage Change-2 Labels | Percentage Change-3 Labels |
|---|---|---|---|
| | | 2 Years | |
| | F1 Score-Macro | F1 Score-Macro | F1 Score-Macro |
| **Naïve Bayes** | 0.58 | **0.59** | 0.41 |
| Logistic Regression | 0.52 | 0.53 | 0.32 |
| Random Forest | 0.53 | 0.56 | 0.37 |
| Gradient Boosting | 0.37 | 0.19 | 0.1 |

On comparison of the results, it was evident that cross-validation did not aid in increasing the F1 score for any of the significant model. There was a minimal increase in the F1 scores. Naïve Bayes continuously achieved the highest score among the other models. On the other hand, Gradient boosting regressor model performance kept degrading and with Percentage change three labels it had given the lowest scores.

The FinALBERT model was also fine-tuned with FinancialPhraseBank dataset. This dataset was also used in FinBERT for testing the performance of their model. We have experimented on sentences with 100% agreement. It consists of 2264 English sentences selected from financial news. These sentences are manually annotated by 16 professionals from the business and finance world. Two configurations were used to fine-tune the model. Two labels (Positive and Negative) and three labels (Positive, Neutral, and Negative) were used. Table 8 depicts the results achieved.

**Table 8.** Results of fine-tuning FinALBERT with FinancialPhraseBank dataset.

| Labels | FinancialPhraseBank – AllAgree | |
|---|---|---|
| | F1 Score-Macro | Accuracy |
| **2 labels** | 0.83 | 84% |
| **3 Labels** | 0.80 | 85% |

Both the label types attained similar F1 macro average score. Unlike our labelling methods, the labels in FinancialPhraseBank are based on sentiment due to which the difference in result is definitive.

To complete the study, the last step was to make a prediction based on proposed FinALBERT model. As the FAANG dataset for one year is fine-tuned with FinALBERT, the dataset was divided in 90:10 ratio for training and testing. For prediction, the two weeks of data for each company is chosen. Predictions are made solely on percentage change labelling method with two labels, i.e., Positive and Negative. Firstly, the labels are predicted using the fine-tuning; these labels are then compared with the actual labels that act as a gold standard for the prediction process. Figure 6 shown below displays predicted labels for AAPL (Apple) StockTwits messages of two weeks.

| | symbol | message | Date | label | pred_label |
|---|---|---|---|---|---|
| 0 | AAPL | peak profit last 6 expired option alerts aapl ... | 2020-07-19 | 1 | 1 |
| 1 | AAPL | aapl jul 17 382 50 calls option volume 144 44 ... | 2020-07-19 | 1 | 1 |
| 2 | AAPL | tsla market true bubble territory profitable c... | 2020-07-19 | 1 | 1 |
| 3 | AAPL | aapl analyzed 26 analysts buy consensus 86 ana... | 2020-07-19 | 1 | 1 |
| 4 | AAPL | aapl new article dogs dow august 4 adopt ignore | 2020-07-19 | 1 | 1 |
| ... | ... | ... | ... | ... | ... |
| 4173 | AAPL | sharing trading secrets video published using ... | 2020-07-14 | 1 | 1 |

**Figure 6.** True labels vs predicted labels.

In the above image "pred_label" is the labels which are predicted ad "label" is percentage change labels. Both labels are used to calculate precision for each label class. It is a fraction of positives predicted, which are positive. For our study, we are taking the threshold of 0.75 for the precision value. Table 9 below shows precision score result.

**Table 9.** Precision score for a positive label class.

| Label | Precision | Recall | F1-score | Support |
|---|---|---|---|---|
| 0 | 0.38 | 0.12 | 0.19 | 955 |
| 1 | 0.73 | 0.92 | 0.81 | 2420 |
| Accuracy | | | 0.69 | 3375 |
| Macro Avg | 0.55 | 0.52 | 0.50 | 3375 |
| Weighted Avg | 0.63 | 0.69 | 0.63 | 3375 |

As evident from the above result, label class 1 (Positive) has less than 0.75 precision value; this predicts stock movement as there are fewer positives for which investment should not be made. Currently, we are displaying a message as "Avoid investing!" and "Invest!" according to the precision value achieved and the decided threshold. For the AAPL company, the prediction is to "Avoid investing" as Predicted precision is 0.726.

## 6. Discussion

The main contribution for this study is a stock movement prediction model which is trained on unsupervised data labelled using the changes in stock prices. A strong correlation exists between the stock price changes and the public sentiments posted as messages on different social media platforms [21]. There are a few models currently available in the market which are based on this correlation. These models check if there are a greater number of positive tweets for a particular day that would mean the stock is performing well and there would be an increase in the price. The main issue in the training of such models is classified as Twitter dataset. In most of them, the tweets are classified (or labelled) by experts or by sentiment analysis tools like VADER, SentiWordNet, TextBlob, etc. The manually tagging of tweets is time-consuming and requires a lot of resources. Our model uses new labelling techniques where the messages are classified into positive, negative, and neutral based on the increase or decrease in the stock price. It makes use of the correlation and works on the assumption that if the price for a stock increased on a particular day then for that stock the tweets would be positive, and thus can be labelled as positive and vice-versa. Apart from this, we proposed a new model FinALBERT, which is ALBERT model trained on financial datasets. Another major limitation of currently available stock movement prediction models available today is that they are missing financial dataset, which would include the financial jargons required for appropriate predictions. Thus, the proposed FinALBERT model is pre-trained using a combination of three datasets (Reuters-21578, which consists of data for money supply or exchanges, AG News which consists of data for the Business category and Bookcorpus which is collection of text taken from Wikipedia) which would help in building a financial specific pre-trained model. We fine-tuned this pre-trained model with different size on the provided StockTwits dataset and experimented with various hyperparameters. For comparison, we experimented with various other traditional machine learning algorithms and transformer-based models.

### 6.1. Evaluation of Labelling Techniques

Comparison between labelling techniques: Out of the three labelling techniques we experimented with, the Percentage change technique with two labels gave the best results on every model. This technique performed better than binary classification labelling technique because it would only classify a tweet as positive if the price increase is greater than 0.5% whereas the binary classification technique simply classifies a tweet as positive even if there is a slight price increase. As we know, a strong correlation exists between the StockTwits messages and stock price changes. In percentage change labelling technique, the threshold for price change is higher. The positively labelled tweets would be more similar to each other, and similarity among negative labelled tweets would be more. Since similarity among two tweets labelled differently would be less, the models can predict the tweets more accurately as compared to when they are trained with data labelled with other labelling techniques.

### 6.2. Model Evaluation

As stated earlier, we experimented with various traditional machine learning algorithms, transformer-based models, and neural network with different word embeddings on all three labelling techniques. Overall, out of all the models experimented with the best results were given by the transformer model BERT and Naïve Bayes algorithm across one year and two years of data for all the labelling techniques.

Among transformer models, BERT model outperforms FinBERT and the proposed FinALBERT model. FinALBERT model did not give the desired results mainly because of the pre-trained model. We used Reuters-21578, AG News and 10% of Book Corpus dataset. The default number of training steps for Albert pre-training equal 1,25,000 [14], but we were able to pre-train the model only on 10,000 steps. As the pre-trained model size was small, we tried improving the model performance by changing hyperparameter values (learning rate, batch sizes, training for more epochs). But even with these changes, the model performance did not improve. We noticed that the FinALBERT model is very sensitive to hyperparameters, and it was initially giving a labelling bias issue because of these parameters. Labelling bias here means that after model training, all the tweets were getting classified as positive (class 1) in the testing dataset. After extensive research, we found that for FinALBERT model to perform well and not have labelling bias, it requires a combination of appropriate training data size and aligned hyperparameters values. Once we aligned the data size and hyperparameter values according to Stanford Sentiment Treebank (SST-2) dataset in ALBERT by [14], we were able to get the desired model results and fix the labelling bias issue. The hyperparameters used for this data in the paper are learning rate as 1.00E-05 and batch size as 32. When we experimented by changing the learning rate to 2.00E-05 or 5.00E-05, we again saw a drop in macro average and labelling bias issue. We also found that to fix labelling bias, we had to add a parameter known as weight decay in the AdamW optimiser. This parameter helps in reducing the overfitting of the model, and the generalisation of the model is improved [31].

While the BERT model outperformed other transformer models because the bert-base-uncased pre-trained model is made up of large Wikipedia data, which consists of approximately 2,500 million words and a book corpus which includes 800 million words; thus, the vocabulary is large and includes common English words. Whereas as mentioned earlier, for the pre-training FinALBERT model, we used a smaller dataset and a smaller number of training steps. Additionally, on further research, we found that ALBERT had outperformed BERT in Stanford Sentiment Treebank (SST-2), a sentiment analysis dataset only when the pre-trained model was albert-xxlarge (not in the case of albert-base). While for pre-training FinALBERT we used the architecture of albert-base as the albert-xxlarge model is computationally very expensive due to its very large structure (large hidden size and more attention heads). We cannot pre-train FinALBERT using albert-xxlarge because of hardware limitations. Thus, the performance of BERT model was more than that of FinALBERT model.

FinBERT performed better than FinALBERT model in all the three labelling techniques. For pre-training, the FinBERT model the TRC2-financial data was added to the bert-base-uncased model. This data consists of about 29 million words and is familiar with financial jargon. Thus, here also the pre-training data was large as compared to our pre-trained FinALBERT model. FinALBERT, when fine-tuned with FinancialPhraseBank dataset gives exceptional results. It consists of 2264 English sentences selected from Financial news. These sentences are manually annotated by 16 professionals from the business and finance world. The gold standard labels in FinancialPhraseBank are based on sentiments of each financial sentence, which helped us infer that FinALBERT can perform well with sentiment analysis as it contains financial domain vocabulary along with generic English vocab. The results outperform when compared to the stock price-based labels for the same reason.

Among the traditional models, the Naïve Bayes model outperforms all other models as it performs well for the large dataset and works well with the categorical dataset. It works well with the large dataset because every feature is considered independently, and probability is calculated for individual categories before predicting the one with the highest. Additionally, the time taken for training the Naïve Bayes model was drastically less when compared to transformer models.

In the traditional models, TF-IDF vectoriser performed better than count vectoriser. To test this, we implement the traditional models using Grid Search using k-fold cross-validation. For all the models, with 5-fold cross-validation Grid Search gave the best parameter as use TF-IDF from the pipeline function. Additionally, TF-IDF vectoriser assigns weight to a word based on the count of the word in a document (term frequency) with its inverse frequency which means its occurrence across all the documents (tweets in our case) while count vectoriser assigns weight by simply counting the words (Stop words like "the", "a" would be given more weightage). Thus, all the traditional models have used TF-IDF for feature extraction.

We also noticed that the model performance did not improve when we used different embedding techniques for feature extraction and training it on neural network models. The neural network-based models like BiLSTM are designed to analyse the sequence of inputs, and our data is not classified based on the sentences but on the price. Because of this reason, these models did not perform well. Additionally, because of the same reason, the attention layer over the BiLSTM was not able to improve the performance. Among all the experiments BiLSTM with FinALBERT word-embeddings performed considerably well with one exception where Word2Vec with BiLSTM performed the best with percentage change two labels among the neural network-based models.

Overall, we only saw a 1% difference in model performance between Naïve Bayes the traditional machine learning model and the transformer-based model BERT. However, as shown in the figure below while comparing the average training times for all models, Naïve Bayes takes approximately 5 minutes to train, while the BERT model, on the other hand, takes 9 hours as transformer-based models have large architectures and are computationally expensive. Thus, as our labelling techniques are based on stock price and not the sentiments of a sentence (or sequence of the sentence), we believe that the implementation of transformer models on this labelled dataset did not help in improving the model performance. Training time for different models are compared and shown in Figure 7. BERT has the highest computation time whereas Naïve bayes has the lowest.

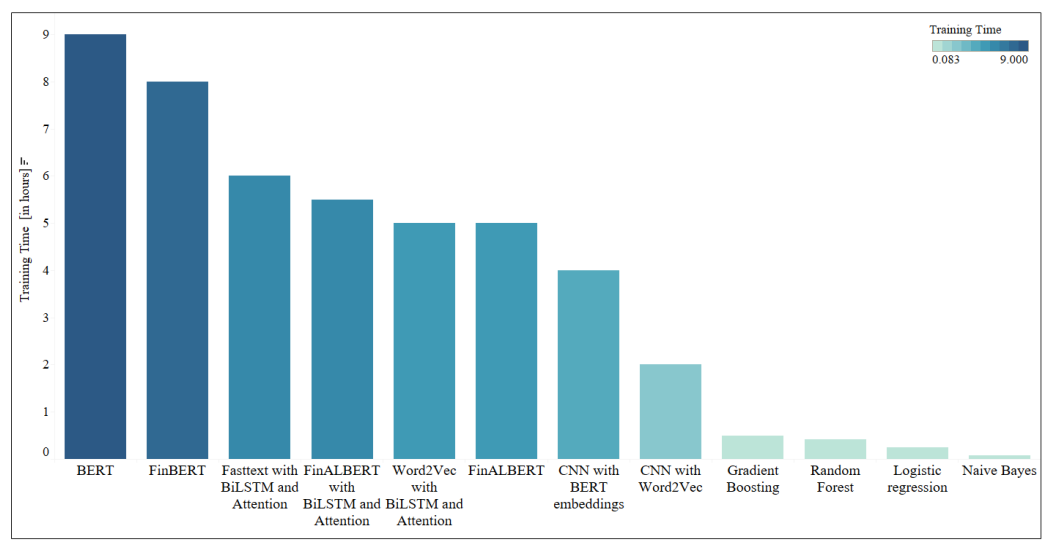

**Figure 7.** Training time comparison.

## 7. Limitation and Future Work

### 7.1. Limitations

#### 7.1.1. External Datasets

The financial datasets currently available in the market are limited or are not openly available [31]. Thus, we had to work with alternative datasets for pre-training. The datasets used are 10% of bookcorpus dataset for the building the English vocabulary, for building financial vocabulary Reuters-21578 and business news filtered form AG_News which provided useful English vocabulary but does not contain specific financial jargons. Additionally, pre-training of FinALBERT model was done on a smaller dataset because of hardware limitations.

#### 7.1.2. Training Time

The training time of transformer-based models is extremely high because these models have large architectures and are computationally expensive. With only a slight increase in model performance and such a high computational time, it was not beneficial to implement these models as compared to traditional machine learning models.

#### 7.1.3. Unlabelled Data

Since the transformer-based models work well in getting the sentiments of the text, and our labelling technique does not consider sentiments of the tweets, the performance of state-of-the-art models is considerably lower. We labelled the tweets by the stock price change and most of the research of the state-of-the-art models done on the basis on sentiment-based labelling due to which we were not able to compare the performance of FinALBERT model with state-of-the-art models properly.

7.1.4. Imbalanced Label Classes

The dataset used for fine-tuning the FinALBERT model with percentage change three labels had a bias data. The number of neutral labels was considerably lower than the number of positive and negative, which also affected the accuracy of FinALBERT while executing with three labels. There was also a slight difference in the number of positive and negative labels, though the difference was meagre.

*7.2. Future work*

Currently, the labelling of unsupervised data for a stock has been done based on the price changes compared to the previous day. Our work can be further extended by labelling the StockTwits by comparing the price change from the last week. Since major events are talked about for a longer period, considering a week's data captures the public opinions related to the event well which in turn would result in better classification of sentiments.

Regarding the performance of FinALBERT model, it can be optimised by using larger data while pre-training the model as it would increase the vocab and use data that contains more financial vocabulary like TRC2 dataset. Also, pre-training the model with a large number of training steps would thereby improve performance. Fine-tuning of FinALBERT model can be done by increasing the number of epochs as the dataset size, while fine-tuning is large and increasing the number of epochs could also increase the accuracy.

As of now, the FinALBERT model has been pre-trained using albert-base-v2 architecture. As mentioned earlier, the performance of albert-xxlarge architecture is better than bert-large for sentiment dataset. Thus, pre-training of FinALBERT using the albert-xxlarge architecture would improve the performance of the model.

Additionally, the FinALBERT model can be directly used for predicting stocks based on the labels predicted. Since labels predicted from FinALBERT alone are not a good indicator, to understand the sentiments model can be fine-tuned with the dataset labelled based on sentiments along with the changes in stock prices for accurate prediction of stock market movements.

Lastly, traditional machine learning methods and transformer-based can be used in combination with an Ensemble-based method [32]. The ensemble method would assign weights to prediction result based on the accuracy of individual models and give the best results out of them.

## 8. Conclusion

In conclusion, the labelling technique used to classify the messages is based on the changes in stock prices and not on sentiments. Currently, not many experiments are performed on this labelling technique. Among the labelling techniques used the Percentage change technique with two labels gave the best results on every model. FinALBERT model is sensitive to the hyperparameter values and size of the dataset. Even though the model was pre-trained on a considerably smaller dataset, the performance was average compared to the other models. The FinALBERT models need to be pre-trained on a large dataset with increased training steps that were not possible due to hardware limitations to improve the model performance. Compared to traditional models, the transformer-based model's training time was too high, and there was no significant improvement in model performance.

**Author Contributions:** Conceptualisation, M.J., U.N., M.K., P.M. and S.N.; methodology, M.J., P.M. and S.N.; analysis, M.J., P.M. and S.N.; data curation, M.J., P.M. and S.N.; writing—original draft preparation, M.J., P.M. and S.N.; writing—review and editing, M.J., U.N., M.K., P.M. and S.N.; visualisation, S.N. Supervision: M.K. All authors have read and agreed to the published version of the manuscript.

**Funding:** This research received no external funding.

**Institutional Review Board Statement:** Not applicable.

**Informed Consent Statement:** Not applicable.

**Data Availability Statement:** The code and data are available from https://mkhushi.github.io/

**Acknowledgments:** The authors would like to thank Hirvita Kharche and Nidhi Shetty for providing a valuable contribution. Jadurshan Puhalendran collected the stocktwits data.

**Conflicts of Interest:** The authors declare no conflict of interest.